\documentclass[osajnl,a4paper,twocolumn]{revtex4}  
\usepackage{hyperref}

\usepackage[]{graphicx}

\begin{document}

\title [Detection of the tagged or untagged photons in acousto-optic imaging ...]{Detection of the tagged or untagged photons in acousto-optic
imaging of thick highly scattering media by photorefractive
adaptive holography }

\author{ M. Gross$^1$, M. Lesaffre$^{1,2}$, F. Ramaz$^2$, P.
Delaye$^3$, G. Roosen$^3$ and A.C. Boccara$^2$}

\address{$^1$ Laboratoire Kastler-Brossel, UMR 8552 (ENS, CNRS, UMPC), Ecole Normale Supérieure, 10 rue Lhomond 75231 Paris cedex 05 France }

\address{$^2$ Laboratoire d'Optique, Ecole Supérieure de Physique et de Chimie
Industrielles de la Ville de Paris, CNRS UPRA0005, Université
Pierre et Marie Curie, 10, rue Vauquelin, 75231 Paris cedex 05,
France }

\address{$^3$ Laboratoire Charles Fabry de l'Institut d'Optique,
CNRS, Universit\'e Paris-Sud, Campus Polytechnique, RD128, 91127
Palaiseau cedex, France.}

\date{\today}

\begin{abstract}
We propose an original adaptive wavefront holographic setup based
on the photorefractive effect (PR), to make real-time measurements
of acousto-optic signals in thick scattering media, with a high
flux collection at high rates for breast tumor detection. We
describe here our present state of art and understanding on the
problem of breast imaging with PR detection of the acousto-optic
signal.
\end{abstract}

\maketitle \textit{ \noindent{\it PACS\/}: 42.25.Dd Wave
propagation in random media, 42.30.Ms Speckle and moiré patterns,
42.40.Ht Hologram recording and readout methods, 42.70.Ln
Holographic recording materials; optical storage media, 42.40.Kw
Holographic interferometry; other holographic techniques, 42.79.Jq
Acousto-optical devices\ }

\bigskip



%
%

\section{Introduction}

The field of acousto-optic imaging has been strongly stimulated by
the deep and complete paper of W. Leutz and G. Maret
\cite{Maret95}. In this paper the authors give a very clear view
of the tricky interactions between light and sound in random
media; this is why this work has stimulated a new active field now
more "biomedical imaging" oriented and new detection schemes.\\

The present paper  is at the frontier of two physical domains that
are \begin{enumerate}
    \item {detection of weak light signal by using
photorefractive crystals,}
    \item {breast cancer imaging by
detection of the ultrasonic modulation of the light scattered
through the breast.}
\end{enumerate}  Here, our purpose is to make a brief review of
these two domains, and to describe the photorefractive detection
of the scattered light modulated component in a pedagogical
manner. By the way, we will describe our present state of art and
understanding on the problem.

In this paper, we will first describe the basics principle of
ultrasonic modulation of light imaging. We will, in particular,
introduce the concept of "ultrasonic tagged photons", which
represents the weak signal to be detected. We will then describe
how photorefractive adaptive holography can be used to detect the
tagged photons. One must notice that all the groups working on the
subject, exempt to us, do not detect the "tagged" photons, but
"untagged" ones. The tagged photon signal is measured indirectly,
since the total number of scattered photons (tagged + untagged)
does not depend on the ultrasound. We will describe our technique
and present our experimental results. In all these descriptions,
we must not forget one difficulty that results from the
decorrelation of the light that travels through breast organ. This
effect known as speckle decorrelation is both due to the brownian
motion of the scatterers, and to the breast inner motions (blood
flow ...). In a typical in vivo situation, with $4cm$ breast
thickness, the "speckle decorrelation time" is in the $0.1$ to
$1ms$ range. It is thus necessary to match the so called
"photorefractive response time" with the "speckle decorrelation
time". This effect, which is huge in breast, is not present in
most of the ultrasonic modulation test experiments, which are
performed with breast phantoms like dead tissues or diffusing
gels. Since the decorrelation affects considerably the detection
sensitivity, it is quite difficult to evaluate the figure of merit
of the different techniques that are proposed to perform breast
imaging. We will see that our setup, which is able to detect both
the tagged and untagged signal, is also able to measure the
photorefractive time \emph{in situ}, \emph{i.e.} with the same
setup, same laser powers and same sample geometry than for breast
imaging experiments. To our knowledge nobody is presently able to
perform ultrasonic modulation imaging though $4cm$ of breast
tissues in vivo. Experiments are under progress and we hope to be
able to reach this aim in a near future.

\section{Acousto-optic imaging}

The combination of light and ultrasound to measure local optical
properties through thick and highly scattering media is a
tantalizing approach for \emph{in vivo} imaging. It is  an
alternative solution to pure optical techniques  for breast cancer
detection. The use of light is motivated by its relative low
absorption in the so called "optical therapeutic window" ($700nm$
to $1000nm$), and by the existence of optical contrasts between
healthy and tumorous areas in this region of the spectrum.

Light is highly scattered within biological tissues, making direct
optical study of thick sample very difficult to perform. Light
scattering is characterized by two length parameters, \emph{e.g
}the scattering length $l_s$, and the light transport mean free
path $l_s^*$. The scattering length $l_s$ characterizes the memory
of optical phase, and corresponds to the average distance that
separates two scattering events. The light transport mean free
path $l_s^*$ characterizes the memory of the light propagation
direction. In tissues, $l_s$ is typically $50$ to $100~ \mu m$,
while $l^*_s$ is $10\times$ larger ($0.5$ to $1mm$). Absorption of
light is characterized by the absorption length $l_a$, which is in
the 1 cm to 10 cm range. Absorption strongly depends on the nature
of the tissue (optical contrast).

Because of scattering, direct imaging cannot  be performed through
more than a few millimeter thick samples. Contrarily to light,
ultrasound (US) beams are ballistic in biological tissues. US
gives thus access to millimeter range spatial resolution in thick
sample (up to $4cm$) yielding the development of the acousto-optic
imaging that combines optics and ultrasound
\cite{Wang95b,Kempe97}.

\subsection{Principle: the tagged photons}

\begin{figure}[]
\begin{center}
\includegraphics[width =8.2 cm,keepaspectratio=true]{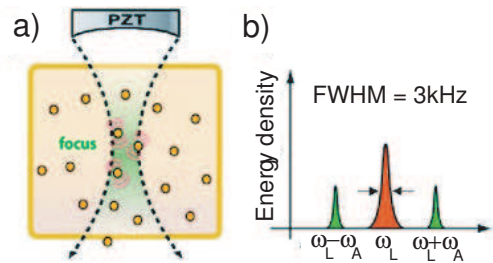}
\caption{Principle of acousto-optic imaging. a) Motion of the
scatterer at $\omega_{A}$. b) Spectrum of the diffused light:
carrier ($\omega_L$), and sideband i.e. tagged photons ($\omega_L
\pm \omega_A$).} \label{fig_acousto_optic}
\end{center}
\end{figure}

Acousto-optic imaging is  a hybrid technique, which combines,
thanks to the acousto-optic effect, ultrasound and light. US are
applied in the region of interest, within the thick scattering
sample (see Fig.\ref{fig_acousto_optic} a). They make the
scatterers vibrate. A CW laser (frequency $\omega_L$) illuminates
the sample. The vibration of the scatterers at the acoustic US
frequency $\omega_{A}$ ($2$ MHz typically) modulates the phase of
the photons that are scattered by the sample. This is the
so-called acousto-optic effect.

The light exiting the sample contains thus different frequency
components (see Fig.\ref{fig_acousto_optic} b). The main component
(the carrier) is centered at the laser frequency $\omega_L$. It is
related to the diffused photons, that do not interact with the US.
The sideband components are shifted by the US frequency $\omega_L
\pm \omega_{A}$. The  sideband photons, which result from the
interaction between light and US, are called "tagged photons"
(\emph{i.e} photons tagged by the US).

The weight of the tagged photons components depends on the optical
absorption in the region of interest, where the US beam is
focused. Acousto-optics imaging stands in detecting selectively
the tagged photons. An image of the sample optical absorption can
be then built-up in scanning the US over the sample. Note that one
of the difficulty in living tissues results from the motion of
scatterers (\emph{e.g} blood flow) which broaden the carrier and
sideband lines (see Fig.\ref{fig_acousto_optic} b). \emph{In vivo}
measurements through 4 cm breast tissues yield a broadening of
$1.5 kHz$ (Full Width at Half Maximum: FWHM)
\cite{Gross_2005,lev2003}.

\subsection{State of the art for the detection of  the tagged
photons } Many techniques have been proposed to detect the tagged
photons. Marks et al. \cite{Marks93} investigated modulation of
light in homogeneous scattering media with pulsed ultrasound. Wang
et al. \cite{Wang95,Wang95b} performed ultrasound modulated
optical tomography in scattering media. Lev et al. studied
scattering media in the reflection configuration \cite{lev2000}.
Wang and Shen \cite{Wang98b} developed a frequency chirp technique
to obtain scalable imaging resolution along the ultrasonic axis by
use of a one-dimensional (1D) Fourier transform. Lev et al. use a
set of optical fibers coupled to a single photo-detector
\cite{lev2000,lev2002,lev2003} that allows to work with samples,
which decorrelate in time. Leveque et al.
\cite{Leveque99,Leveque00,Leveque01} performed parallel detection
of multiple speckles on a video camera and demonstrated
improvement of the detection signal-to-noise ratio of 1D images of
biological tissues. The parallel detection has been still improved
by Gross et al., which performs holographic detection reaching the
shot noise sensitivity limit \cite{Gross_03}, and by Atlan et al.,
which get resolution on the US propagation axis by using an
holographic pulsed technique \cite{Atlan_2005}.

All these methods exhibit two main limitations. First, the
\textsl{optical etendue} (defined as the product of the detector
area by the detector acceptance solid angle) of the detection
system is not optimum, since it is much lower than the
\textsl{etendue} of the tagged photons source. This
\textsl{etendue} is the area of the sample (several cm$^2$)
$\times$ the emitting solid angle (which is about $2\pi$ since the
light is diffused by the sample in all direction). With a mono
detector (photodiode) \cite{lev2000,lev2002,lev2003} the detection
\textsl{etendue} is about $\lambda^2$. With a multi detector like
a CCD camera
\cite{Leveque99,Leveque00,Leveque01,Gross_03,Atlan_2005} the
\textsl{etendue} is $N\lambda^2$, where $N$ is the is CCD number
of pixel ($N \sim 10^6$). Even with a camera, the \textsl{etendue}
of detection is about $\times 1000$ lower than the
\textsl{etendue} of the emission.

The second problem occurs within living sample: the scatterers
move, yielding in the frequency space a broadening of the tagged
photons spectrum, as shown on Fig.\ref{fig_acousto_optic}b
\cite{Gross_2005,lev2003}. This effect corresponds, in the time
space, to a decorrelation of the tagged photons speckle pattern.
Since all the methods described above perform coherent detection,
the bandwidth of detection is limited by the detector bandwidth.
With camera (there is no problem of bandwidth with photodiode, but
the \textsl{etendue} is much lower), the bandwidth is roughly
equal to the camera image frequency $\omega_{CCD}$, which is in
general much lower ($\omega_{CCD} \sim 10...100$ Hz) than the
tissue broadening (3 kHz). It is still possible to work with fast
camera (kHz), but in that case i) the camera quantum efficiency is
lower (CMOS), and ii) the number of pixel $N$ is limited, because
$N \times \omega_{CCD}$ is the flux of information to transfer to
the computer, and this flux is limited ($< 10^6 ... 10^7 s^{-1}$).

\section{The photorefractive (PR) detection of the acousto optic signal}

\begin{figure}[]
\begin{center}
\includegraphics[width =8.2 cm,keepaspectratio=true]{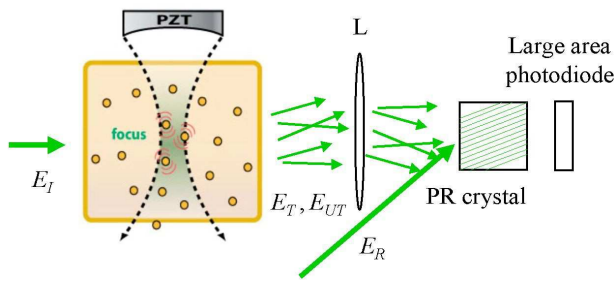}
\caption{Principle of the PR detection of the acousto-optic
signal. PZT: ultrasound transductor, $E_I$: illumination optical
field; $E_{T}$,$E_{UT}$ tagged or untagged field scattered by the
sample; PR crystal: photorefractive crystal;  L: lens that collect
the scattered field into the crystal and photodiode; $E_R$: PR
crystal reference (or pump) field. } \label{fig_FIG_PR_detection}
\end{center}
\end{figure}

More recently has appeared a new tagged photons detection
technique that is based on the photorefractive effect (PR) and
that is illustrated by Fig.\ref{fig_FIG_PR_detection}. The light
that is scattered by the sample ($E_{T}$ or $E_{UT}$ for the
tagged or untagged field) is detected by a photorefractive
detector (PR crystal + photodiode PD) that is pumped by a
reference field $E_R$.

Since the crystal and the photodiode might be quite large (up to
$1 cm^2$) and since the light is collected by a large Numerical
Aperture (N.A. $\sim 1$) collecting lens, the photorefractive
detection benefits of a high \emph{etendue}, about $100 \times$
larger than in a typical camera with $N \sim 10^6$ pixels. We will
see  that the detection bandwidth is the inverse of the
"photorefractive time" $T_{PR}$.  We get for example, for a
$1W/cm^{2}$ pump beam, $1/T_{PR}\sim$ 1 kHz
\cite{Ramaz_04,Murray_04}. This bandwidth, which is about $100
\times $ larger than for a $N\sim 10^6$ CCD camera \footnote{we
implicitly exclude here fast CMOS camera because of poor  quantum
efficiency and noise, and because of finite bandwidth for the data
transfert form camera to computer}, is within the range of the
linewidth  ( $\sim 3$ kHz) of the light scattered  \emph{in vivo}
by a breast organ \cite{Gross_2005}.

\subsection{The volume hologram}

Photorefractive effect arises in materials that present both
electrooptic effect and photoconductivity, which combination
allows to transform a non uniform illumination of the material
into a spatial variation of the refractive index \cite{Yariv}.
When illuminated by the interference pattern between an object and
a reference beam, the material records a hologram, \emph{i.e} the
amplitude and phase of the object beam. This hologram is dynamic
meaning that it can follow the interference pattern fluctuations
slower than the response time $T_{PR}$ of the material, also
meaning that only slowly moving hologram are recorded.

\begin{figure}[]
\begin{center}
\includegraphics[width =8.2 cm,keepaspectratio=true]{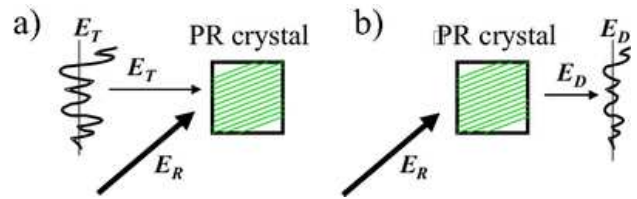}
\caption{Principle of holography using the photorefractive
effect.} \label{fig_PR_effect}
\end{center}
\end{figure}

\begin{figure}[]
\begin{center}
\includegraphics[width =8.2 cm,keepaspectratio=true]{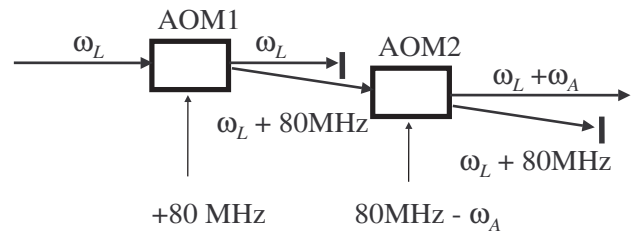}
\caption{Method for tuning the frequency $\omega_R$ of the
reference beam:  $\omega_R=\omega_L+\omega_A$ for example. AOM1
and AOM2 are acousto optic modulators.} \label{AOM.eps}
\end{center}
\end{figure}

The first effect is the recording of the signal beam information
(phase and amplitude of the signal field $E_T$) within the PR
crystal under the form of local changes of the crystal refractive
index $n$ yielding a volume hologram (Fig.\ref{fig_PR_effect}a).
Since the recording takes a finite time $T_{PR}$ ($0.1$ to $10ms$
in our device), the PR effect selects, within the signal beam, the
components whose frequency is close or equal to the reference beam
frequency $\omega_R$.

Here, the large, almost flat field, reference beam (field $E_R$)
interferes with signal field $E_T$ onto the PR crystal. In order
to select the tagged or untagged photons, the frequency $\omega_R$
of the reference beam is made equal to the tagged or untagged
photons frequency: $\omega_R=\omega_L\pm\omega_A$ or
$\omega_R=\omega_L$ respectively. To adjust $\omega_R$ one can use
for example two acousto-optic modulators AOM1 and AOM2 (Bragg
cells) as shown on Fig.\ref{AOM.eps}. With such a choice, the
interference pattern $E_T E^*_R$ of the reference beam with the
selected photons beam varies slowly in time. The selected beam
information can thus be grooved within the PR crystal volume
hologram.

In the case of a perfect monochromatic signal beam,  the local
variation of the hologram refractive index $\delta n$ is simply
proportional to the modulation depth of the interference pattern
$(E_T E^*_R)/ (|E_T|^2 + |E_R|^2)$. If a time modulation is added
on the signal (e.g. amplitude or phase modulation), we have to
take into account the finite time $T_{PR}$ needed to groove the
hologram, and we get \cite{Delaye_photoref_1,Delaye_photoref_2}:
\begin{equation}\label{Eq_delta_n}
    \delta n \propto \frac{ \left< E_T E^*_R \right>_{T_{PR}}}{|E_T|^2 +
|E_R|^2}
\end{equation}
where $ \left< ~ \right>_{T_{PR}}$ is the average over the
grooving time $T_{PR}$, average which is defined by:
\begin{equation}\label{Eq_delta_n}
 \left< A \right>_{T_{PR}}= \frac{1}{T_{PR}} \int^\infty _0 A(t-\tau)~
 e^{-\tau/T_{PR}}~d\tau
\end{equation}

\subsection{The diffracted beam $E_D$}

The second effect is illustrated by Fig.\ref{fig_PR_effect}b. The
reference beam ($E_R$) is diffracted by  the volume hologram
yielding a diffracted  beam ($E_D$). The diffracted field $E_D$ is
simply proportional to the hologram refractive index changes
$\delta n$ and to the reference beam field $E_R$. We get thus:
\begin{equation}\label{Eq_E_D}
    E_D \propto \frac{ \left< E_T E^*_R \right>_{T_{PR}}}{|E_T|^2 +
|E_R|^2} E_R
\end{equation}
In typical application the reference beam intensity is much larger
than the signal beam one, and except of the average over $T_{PR}$,
$E_R$ and $E^*_R$ simplifies in Eq.\ref{Eq_delta_n} yielding
$E_D\propto E_T$ i.e.
\begin{equation}\label{Eq_delta_n}
    E_D \simeq \eta E_T
\end{equation}
where $\eta=0.1 .. 0.5$ is a numerical factor which mainly depends
on the crystal.

Eq.\ref{Eq_delta_n} is valid, when the decorrelation of the signal
field $E_T$ can be neglected during the grooving time $T_{PR}$,
i.e. when
\begin{equation}\label{Eq_decorrelaton_cond}
    \delta \omega T_{PR} \ll 1
\end{equation}
where $\delta \omega$ is the frequency width of the signal beam
($\Delta \omega \sim 3$ kHz for the breast).

We have to notice that an increase of the reference beam intensity
$|E_R|^2$ does not change $\eta$, but reduces the grooving time
$T_{PR}$, since $T_{PR} \propto 1/|E_R|^2$. The main advantage of
increasing the reference beam power $|E_R|^2$  is thus to reduce
$T_{PR}$ enough to neglect the signal field decorrelation.
Condition of Eq.\ref{Eq_decorrelaton_cond} is then fulfilled, and
the Eq.\ref{Eq_delta_n} limit can be reached.

Since the volume hologram has recorded the mode structure of the
signal beam versus reference beam interference pattern, and since
the pump beam is diffracted by the hologram, the diffracted beam
($E_D$) has the same mode structure than the signal beam ($E_T)$
(see  Eq.\ref{Eq_delta_n}). This result is illustrated by
Fig.\ref{fig_PR_effect}b where   $E_D$ is displayed with the same
shape than $E_T$ on Fig.\ref{fig_PR_effect}a, but with a smaller
amplitude ($\eta <1$).

The signal ($E_T$) and diffracted ($E_D$) beams  are thus
spatially coherent. They can interfere constructively (or
destructively) on a large area ($\sim 1 ~\textrm{cm}^2$) light
mono detector (i.e. a photodiode). This property will be useful to
detect efficiently the tagged and untagged photons signal.

\subsection{Detection of the tagged photons}

\begin{figure}[]
\begin{center}
\includegraphics[width =8.2 cm,keepaspectratio=true]{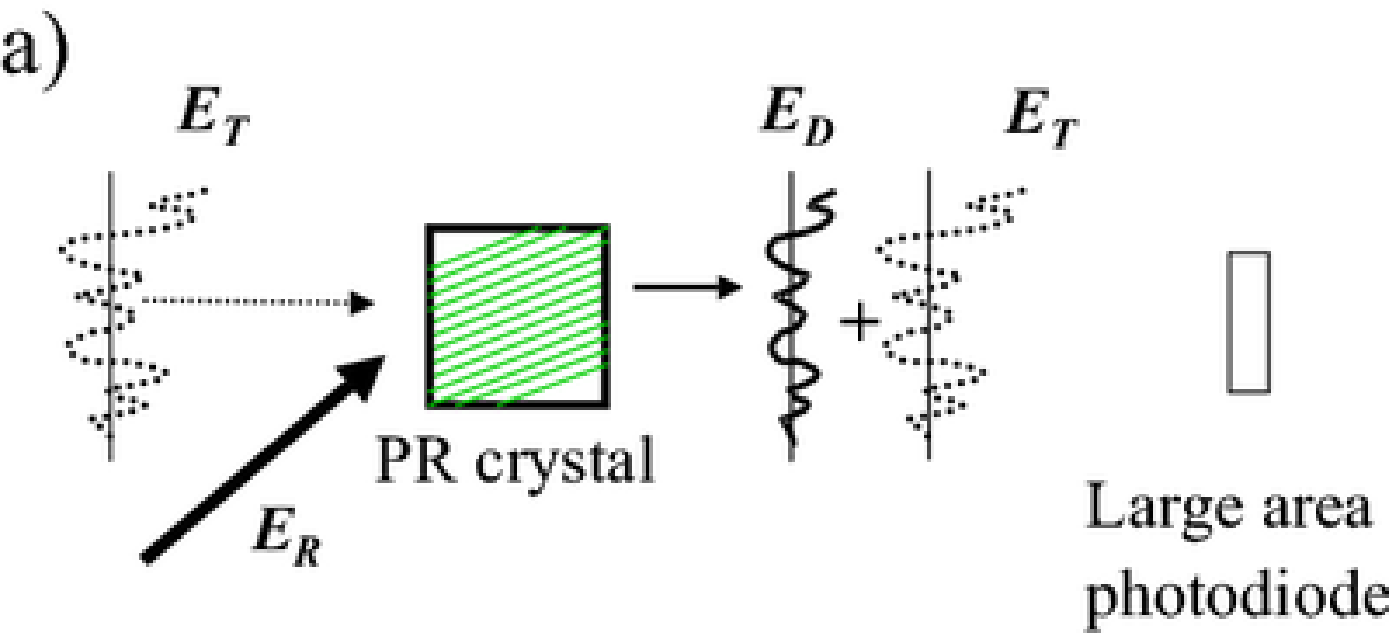}
\includegraphics[width =8.2 cm,keepaspectratio=true]{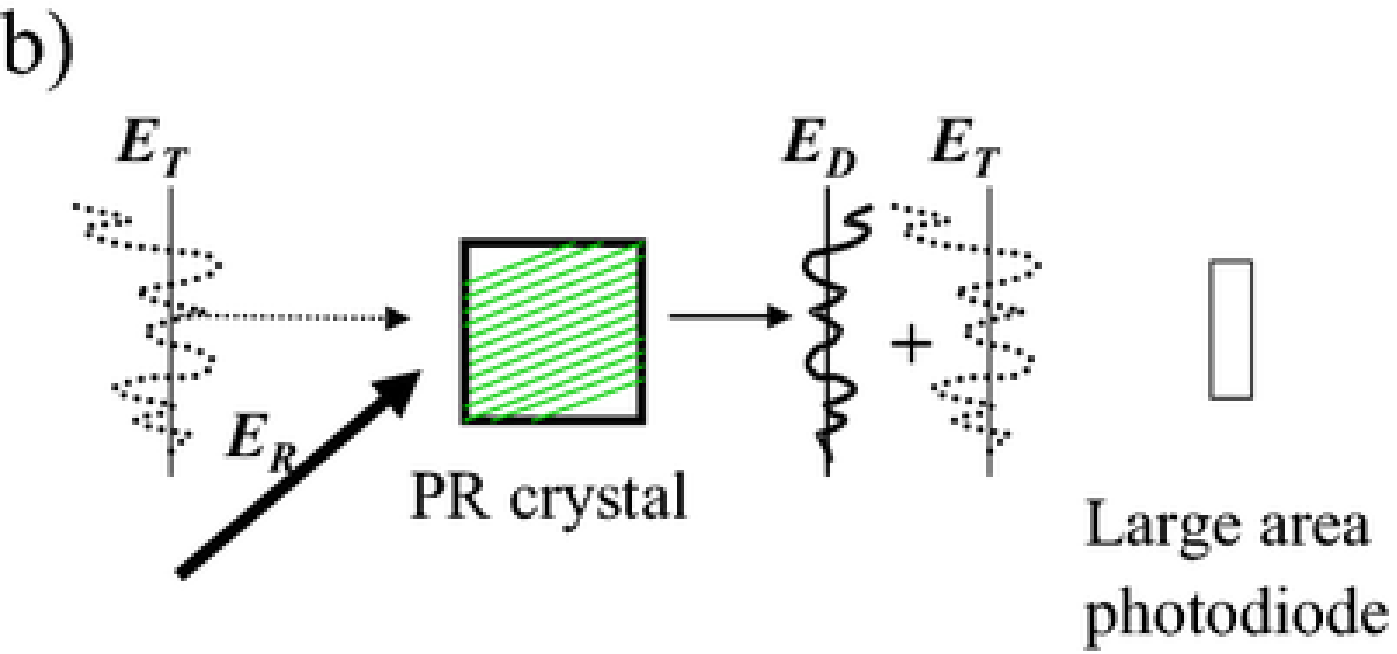}
\includegraphics[width =8.2 cm,keepaspectratio=true]{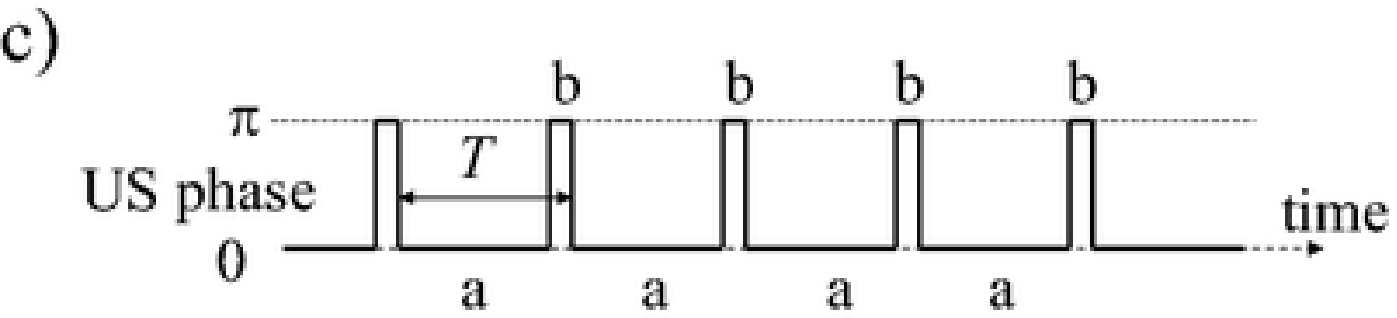}
\caption{a,b) Detection of the tagged photons, when the phase of
US beam is zero (a) and $\pi$ (b). c) Phase of the US beam.  }
\label{fig_detection_tag}
\end{center}
\end{figure}

The principle of tagged photons detection is illustrated by
Fig.\ref{fig_detection_tag}. The phase $\varphi$ of the US beam is
reversed periodically ($\varphi=0$ or $\pi$) with period $T$ (see
Fig.\ref{fig_detection_tag}c). The phase of the tagged photons
field $E_T$, which follows the US phase, is then reversed too.

To simplify the discussion, we will neglect the decorrelation of
the tagged photons field (Eq.\ref{Eq_decorrelaton_cond} is
fulfilled). We will also modulate the phase rapidly (with respect
to $T_{PR}$ i.e. with $T \ll T_{PR}$), keeping $\varphi$ zero most
of the time (see Fig.\ref{fig_detection_tag} c), so that the
hologram can be considered as static and unperturbed by the phase
variation.

In that case, the diffracted field $E_D$ will remain nearly
constant: $E_D \simeq \eta \left <E_T \right>$. When $\varphi$ is
zero, $E_T$ and $E_D$ are in phase, they interfere constructively
and the total intensity signal $|E_T+E_D|^2$ is maximum (see
Fig.\ref{fig_detection_tag}a):
\begin{equation}\label{Eq_signal phase +}
    I_{0}=(|E_T+E_D|^2)_{\varphi=0} \simeq |E_T|^2 (1+\eta)^2
\end{equation}
When phase is $\pi$ contrarily, the $E_T$ and $E_D$ are opposite
in phase, and the total intensity signal is minimum (see
Fig.\ref{fig_detection_tag}b):
\begin{equation}\label{Eq_signal phase +}
    I_{\pi}=(|E_T + E_D|^2)_{\varphi=\pi} \simeq |E_T|^2 (1-\eta)^2
\end{equation}

Reversing the phase of the US yields to a modulation of the total
intensity signal equal to :
\begin{equation}\label{Eq_mod_I_varphi}
    I_{0} - I_{\pi} \simeq~ 4~\eta ~|E_T|^2+ ...
\end{equation}

\subsection{Detection of the untagged  photons}

It is a little bit more difficult to illustrate the detection of
the untagged photons by a simple figure, because the calculation
involves to consider both the untagged photons field at the
carrier frequency $\omega_L$, and the tagged photons fields
$E_{T}$ and $E_{T'}$ which evolves at the two sideband frequencies
$\omega_L+\omega_A $ for $E_T$, and $\omega_L -\omega_A $ for
$E_{T'}$. To detect the untagged photons, we tune the reference
beam frequency $\omega_R$ at the untagged photons frequency:
$\omega_R=\omega_L$, and we modulate the US beam intensity by
turning on and off the US beam.

To simplify the discussion, we will neglect again the
decorrelation of the tagged photons field
(Eq.\ref{Eq_decorrelaton_cond} is fulfilled). We will also
modulate the US beam rapidly (with respect to $T_{PR}$ i.e. with a
period $T \ll T_{PR}$). Let us call $E_{U}$ and $E_{U'}$ the
untagged photons fields without, and with US beam.  $E_{T}$ and
$E_{T'}$ are the tagged photons fields with US (theses fields are
zero without US).

Since the energy is conserved, the total number of photons
(carrier + sidebands) does not depend on the US. We get thus:
\begin{equation}\label{Eq_cons_energy}
    |E_{U}|^2 =  |E_{U'}|^2 ~+ ~|E_{T}|^2~+ ~|E_{T'}|^2
\end{equation}

The untagged photons field in presence of the US, \emph{e.g}
$E_{U'}$, is spatially coherent with the one without US,
\emph{e.g} $E_{U}$. According to (Eq.\ref{Eq_cons_energy}), its
magnitude can be expressed as follows :
\begin{equation}\label{Eq_untagged_decrease}
    |E_{U'}| =  |E_{U}|\sqrt{1 ~ - ~ \frac{|E_{T}|^2~+ ~|E_{T'}|^2}{|E_{U}|^2}}
\end{equation}

In practical situation, the efficiency of the acousto optic effect
is low and the energy within the sideband is low ($<1\%$) with
respect to the carrier. This means that the untagged photons field
variation is low: $E_{U}-E_{U'} \ll E_{U}$. Whatever the value of
the cyclic ratio modulation is, one can thus consider that the PR
effect involves $E_{U}$ only. We get:

\begin{equation}\label{Eq_E_D_U}
    E_{D} \simeq ~\eta ~ E_{U}
\end{equation}
When the US is off, the field on the detector is $E_{U} + E_D$ and
the detected intensity signal $I$ is:
\begin{eqnarray}\label{Eq_Ion}
    I= |E_{U} + E_D |^2 ~~~~~~~~~~~~~~~~~\\
    =  |E_{U}|^2 + |E_D |^2 + 2 \eta  |E_{U}|^2
\end{eqnarray}
When the US is on, the field on the detector is $E_{U'} + E_D$ for
the carrier, and $E_T$ and $E_{T'}$ for the two sidebands.  The
intensity signal $I'$ is:
\begin{eqnarray}\label{Eq_Ioff}
    I'= |E_{U'} + E_D |^2 + |E_{T}|^2+ |E_{T'}|^2~~~~~~~~~~~~~~~~~~~~\\
            = |E_{U'}|^2  + |E_D |^2 + \eta (E_{U'}E_{U}^{*}+E_{U}E_{U'}^{*})+ |E_{T}|^2+ |E_{T'}|^2
\end{eqnarray}
Taking into account the energy conservation
(Eq.\ref{Eq_cons_energy}), the spatial coherence of
($E_{U}$,$E_{U'}$) and (Eq.\ref{Eq_untagged_decrease}), we get the
modulation of the detected intensity:
\begin{eqnarray}\label{Eq_Ioff}
    I-I'= 2\eta ~ (|E_{U}|^{2}-|E_{U'}|.|E_{U}|)~~~~~~~~~~~\\
       \simeq 2\eta ~  \frac{|E_T|^2+ |E_{T'}|^2}{2}\simeq 2\eta ~ |E_T|^2
\end{eqnarray}
since the weight of the two sidebands components are approximately
the same: $|E_T|^2 \simeq | E_{T'}|^2$.

By comparing Eq.\ref{Eq_mod_I_varphi} and Eq.\ref{Eq_Ioff}, the
detected signal have the same order of magnitude when detecting
either the tagged or the untagged photons, when we consider the
same acoustical energy.

\subsection{Detecting  tagged or untagged photons ?}

To our knowledge, three groups are working on  acousto optic
imaging with PR detection of the signal. Two of them, the R.A. Roy
\cite{Murray_04,Bossy_2005,Sui_2005,Blonigen_2005} and the L.V.
Wang group \cite{Wang03} detect the untagged photons. We are the
third group \cite{Ramaz_04,Gross_Ramaz_2005} and we detect both
the tagged and untagged photons.

Detection of the untagged photons is simpler since it is not
necessary to shift frequency the reference beam
($\omega_R=\omega_L$). The acousto optic modulators of
Fig.\ref{AOM.eps} are thus not needed. Moreover, it is not
necessary to apply the US beam all the time. Untagged photons
detection is thus well suited to detect very short burst of US
beam able to give information resolved along the US beam
propagation direction \cite{Murray_04}. But short US burst yield a
small signal, and signal is needed to image thick breast \emph{in
vivo}.

The detection of the untagged photons corresponds to a small
change on a large signal (white background detection), while the
detection of the tagged photons, which corresponds to roughly the
same absolute value change, yields contrarily to about $100\%$
change on a small signal (black background detection). Tagged
photon detection is thus expected to give less technical noise.
For example, vibrations on the reference beam mirrors, which
modify the length of the pump beam arm, is expected to yield about
$100 \times$ \footnote{here 100 is the untagged versus tagged
photons field ratio} more noise for the untagged configuration
than for the tagged one.

The tagged photon configuration offers more degrees of freedom for
the detection configuration, because the signal and reference beam
can be modulated whether in phase or amplitude.

Since we do not have tested all the possible detection
configurations, making a complete comparison of tagged and
untagged photons detection schemes is out of the scope of the
present paper. We can simply say that for the configurations we
have presently tested, the Signal to Noise Ratio (SNR) is about
the same in the two cases. Since our purpose is to image breast,
we need to improve the detection sensitivity.  We continue thus to
work with our setup that is able to detect both tagged and
untagged photons, exploring configurations that are expected to
yield better SNR. This work is under progress.

\section{Experimental test}

\subsection{Setup}

\begin{figure}[]
\begin{center}
\includegraphics[width =8.2 cm,keepaspectratio=true]{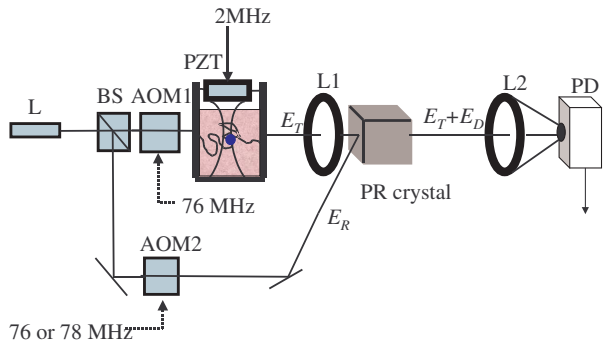}
\caption{Typical experimental setup. L: laser; BS: Beam Splitter;
AOM1, AOM2: acousto optic modulators; PZT: US generator; L1, L2:
light collecting lenses; PR: photorefractive crystal; PD: photo
detector. } \label{fig_setup}
\end{center}
\end{figure}

A typical experimental setup, able to select either the tagged or
the untagged photons, is shown on Fig.\ref{fig_setup}. The main
laser  beam is splitted in an illumination and reference beam by
the beam splitter BS. The US beam ($2MHz$) produced by the
generator PZT is focused within the sample. The frequency offset
of the illumination beam is adjusted by using the two
acousto-optic modulators AOM1 and AOM2 which are excited at
$78MHz$ and $76MHz$ (for selecting the tagged photons), or both
$78MHz$ (for selecting the untagged photons). The light diffused
by the sample is collected by the high $NA\sim1$ (Numerical
Aperture) lenses L1 and L2. L1 collects the light within the PR
crystal that records the hologram of the selected signal beam
(tagged or untagged). L2 collects the interference pattern of the
signal beam ($E_T$) with the diffracted beam ($E_D$) into the
photodetector PD.

In our setup, L is a Nd:YAG laser (1.06 $\mu$m, 1 to 5 W CW
power), the PR crystal is a $1.4\times 1.4\times 2 cm^{3} $ GaAs
crystal \cite{Delaye_94}, and PD is a large area photodiode (0.1
to 0.5 cm$^2$) whose signal is amplified by a transimpedance
amplifier ($R=100 K\Omega$ to 10 M$\Omega$).

In the Murray's setup \cite{Murray_04}, L is a frequency doubled
Nd:YAG (532 nm, 80 mW), the PR crystal is a $5\times 5\times 7
mm^3$ Bi$_{12}$SiO$_{20}$ crystal, whose PR efficiency is improved
by applying a DC electric field, the US frequency is 1.1 MHz, and
PD is an avalanche photodiode. Since Murray detects the untagged
photons, the acousto-optic modulators are not present, but it
should be pointed out that absorption at $532nm$ is more important
than at $1064nm$, and thus it can reduce the thickness of
investigation.

Our setup, which can detect both the tagged and untagged photons,
is expected to be more sensitive, while the Murray's setup, which
is used with short US pulses, is faster.

\subsection{Experimental result}

\begin{figure}[]
\begin{center}
\includegraphics[width =8.2 cm,keepaspectratio=true]{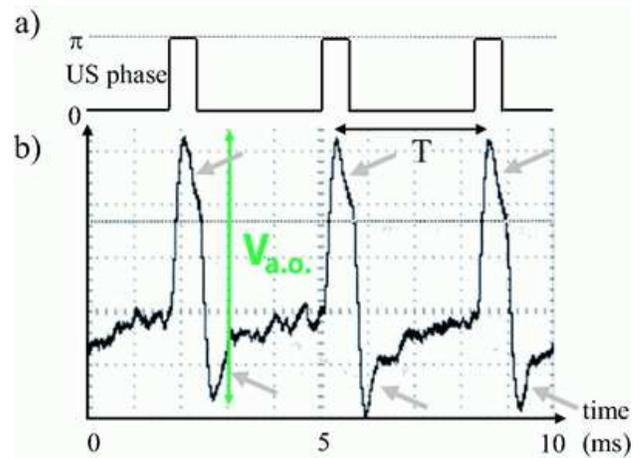}
\caption{Typical tagged photons experimental signal. a) Modulation
of the US phase b) Tagged photons signal. $V_{\textrm{a.o.}}$:
maximum variation of the tagged photons signal. T: phase
modulation period.  Vertical axis is voltage at the output of the
PD amplifier: $100mV$ per $div$.} \label{fig_mod_phase__signal}
\end{center}
\end{figure}

Fig.\ref{fig_mod_phase__signal} shows a typical tagged photons
experimental signal obtained with $0$ to $\pi$ phase modulation of
the US beam. The modulation frequency is $300Hz$ (modulation
period $T=3.33$ ms). The US beam frequency is $2MHz$, with a
maximum US pressure of $2MPa$ at the US beam waist. The main laser
power is $1.2W$. The reference and illumination beam power are
both $300mW$, their areas on crystal and sample are both $1cm^2$.

Measurement is performed with a $4cm$ chicken sample, whose
optical properties (diffusion and absorption) are close to human
breast. As seen, the tagged photons signal SNR is good ($16$ times
averaged). One must notice that the signal is not rectangular like
the phase modulation. In particular, the maximum of signal, which
occurs on the phase plateaus ($\varphi=\pi$ and $0$) is not flat,
but decreases exponentially (see grey arrows on
Fig.\ref{fig_mod_phase__signal}).

This is expected when the PR time $T_{PR}$ becomes shorter than
the phase modulation period $T$. We have measured (see further)
$T_{PR}$ and we have found $T_{PR}=0.5$ ms. Note that $T_{PR}$ can
also be measured on Fig.\ref{fig_mod_phase__signal}, since
$T_{PR}$ is time constant of the grey arrow decay.

This result is very encouraging, because it means that the
detection bandwidth is $1/(2\pi T_{PR})=0.3kHz$. Remember the
signal bandwidth is $\Delta \omega= 1.5kHz$ (HWHM) on the breast.
The detection is thus optimal, within a factor $5$. Since the SNR
is very high (much larger than $5$) in
Fig.\ref{fig_mod_phase__signal}, we expect to get enough SNR to
get significant result with a thick living sample.\

\begin{figure}[]
\begin{center}
\includegraphics[width =8.2 cm,keepaspectratio=true]{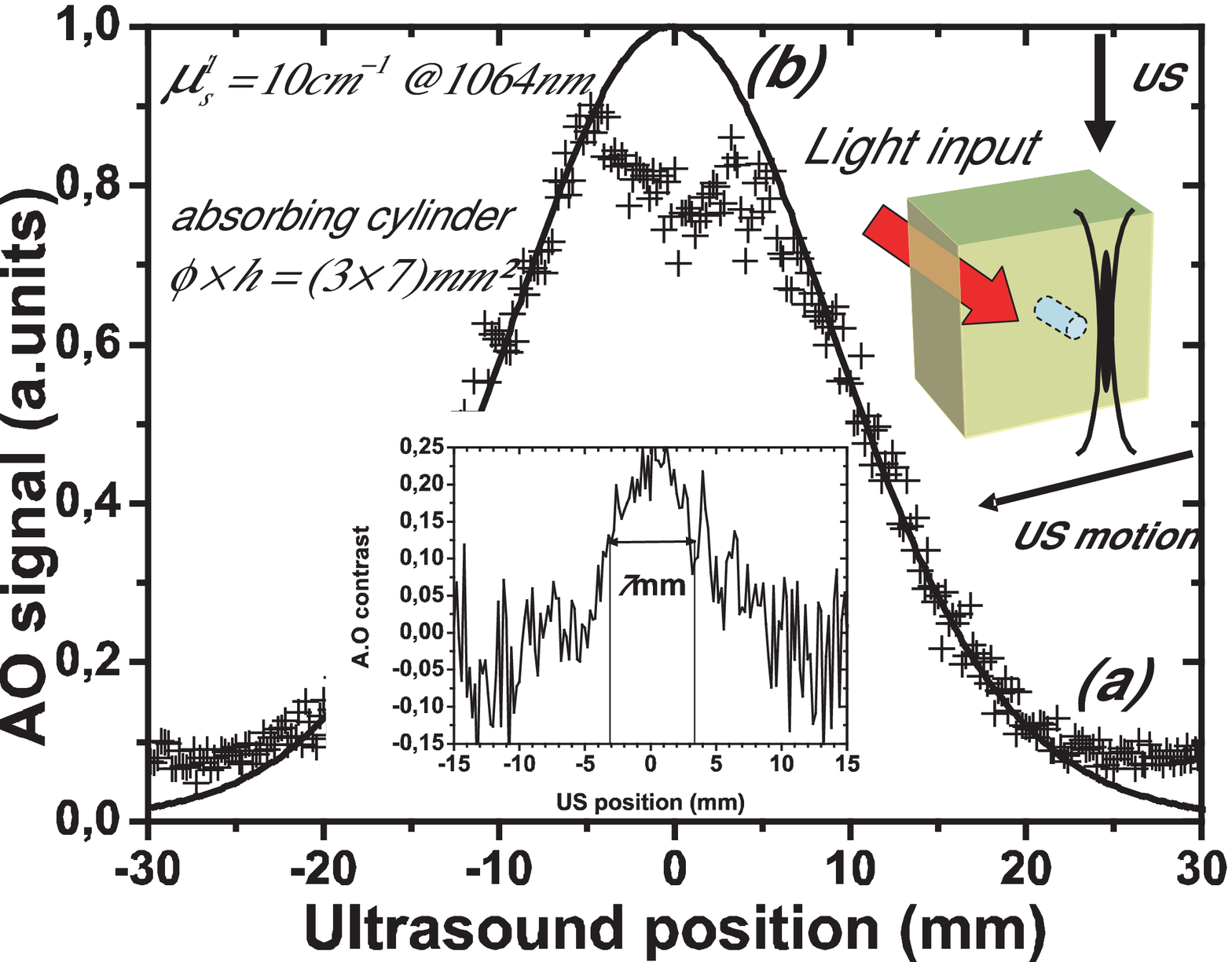}
\caption{Acousto-optic profile (a) with an optical absorber
embedded within an Agar + \emph{Intralipid} phantom of thickness
$t=30mm$ with a reduced scatttering coefficient
$\mu^{'}_{S}=10cm^{-1}$ $@1064nm$ obtained with an anisotropic
(PR) detection configuration. Light input on the sample is
$1W/cm²$ (gaussian illumination $FWHM=1.1mm$). The US pressure
($2.3MHz$,$1.5MPa$) is phase modulated at $3kHz$ with a duty cycle
of $24\%$. Curve (b) represents a fit of the one dimensional
extend of light scattering (e.g $exp(-\mu_{eff}r)/r$ with
$\mu_{eff}=2.2cm^{-1}$) within the US plane (15mm from input
window), weighted by the input illumination
$\mu_{eff}=2.2cm^{-1}$. Inset represents the acousto-optic
contrast $[1-(a)/(b)]$.}\label{fig_profil_inclusion}
\end{center}
\end{figure}

As an other illustration, Fig.\ref{fig_profil_inclusion}(a)
represents a profile of a Agar plus \emph{Intralipid} phantom with
a thickness $t=30mm$ and a reduced scattering coefficient
$\mu^{'}_{S}=1/l^{*}_{s}=10cm^{-1}$ at $1064nm$. The absorption
coefficient of water $@1064nm$ is $\mu_{a}=0.144cm^{-1}$
\cite{Kou_1993}. The sample contains an optical absorber (black
ink), which is a cylinder with a diameter of $3mm$ and a height of
$7mm$ along the laser input direction (perpendicular to the US
beam). The magnitude of the US pressure is approximately of
$1.5MPa$ at $2.3MHz$, with a phase modulation at $3kHz$ and a duty
cycle of $24\%$, corresponding to the maximum of the signal with a
lock-in detection \cite{Ramaz_04,Gross_Ramaz_2005}. The
photorefractive holographic setup is based on an anisotropic
diffraction configuration\cite{Yeh_88}: the reference beam
(\emph{e.g} vertically polarized) diffracts a contribution
(\emph{e.g} tagged-photons field) which is perpendicularly
polarized (\emph{e.g} horizontal); the output speckle from the
sample is $45^{\circ}$-polarized from vertical direction using a
large aperture infrared dichroic polarizer, and a similar analyzer
is positioned in front of the photodetector with an horizontal
polarization axis. Consequently, the speckle and the reference
fields still interfere within the PR crystal in order to build the
hologram, the diffracted reference and the speckle field recombine
onto the analyzer as well. This configuration minimizes the
collection of the unwanted scattered reference by the PR crystal
faces. In this experiment the tagged light is about $\times 10^4$
lower than the total scattered light (untagged photons plus
scattered reference light).

Classically, in the $3D$ diffusion regime and in presence of
absorption, the spatial distribution of energy emitted from a
point source at distance $r$ is given by $\frac{1}{r} ~
e^{-\mu_{eff}r}$, where
$\mu_{eff}=\sqrt{3\mu_{a}(\mu_{a}+\mu^{'}_{s})}$. This effective
parameter indicates that attenuation is increased by scattering,
that lengthens optical pathes.

The continuous envelope Fig.\ref{fig_profil_inclusion}(b)
represents the fit the experimental data
Fig.\ref{fig_profil_inclusion}(a) using this model and taking into
account the gaussian input illumination ($FWHM=1.1mm$). The
effective coefficient $\mu_{eff}$ is founded to be $2.2cm^{-1}$,
close to the theoretical value ($\mu_{eff}=2.1cm^{-1}$) given by
the reduced scattering coefficient of the medium and the
absorption coefficient of pure water at $1064nm$ defined above.
The measured background (around $0.8mV$) corresponds to the noise
of the transimpedance stage of the detection, that is shot-noise
limited at this level of the scattered light. The absorbing
element is revealed by the acousto-optic contrast, \emph{e.g}
$[1-(a)/(b)]$, which is close to $0.22$, and exhibits a
\emph{FWHM} of $7 mm$. This value is connected to the diameter of
the absorbing element ($3mm$), the US resolution (just above
$1.5mm$) and the light transport mean free path $l_s^*$ of the
scattering medium (about $1mm$).

\section{Measurement of the photorefractive time $T_{PR}$}

Most published results on ultrasound light modulation imaging have
been obtained with phantoms, which do not decorrelate in time. In
that case, the PR detection SNR does not strongly depends on the
reference beam power. The power must be large enough to reach the
plateau value for the photorefractive efficiency $\eta$, but
remains low enough  to avoid noise (the reference beam is
scattered by the PR crystal defects yielding a parasitic
photodiode current that brings noise). With phantoms, the best
sensitivity is then obtained with a quite low power reference beam
($<100$ mW in our experiment).

With breast, the light signal is Doppler broadened by the tissues
inner motions (brownian motion, blood flow...) yielding typically
a spectral width of $3kHz$ \cite{Gross_2005}. In order to optimize
the detection efficiency, one must increase the detection
bandwidth $1/T_{PR}$ by increasing the reference beam power to
obtain $1/T_{PR}\sim 3 $ kHz. Optimal detection conditions for
phantoms and breast are thus very different.

To improve the detection sensitivity for future breast experiment,
it is very important to measure $T_{PR}$. To get reliable result,
we have proposed a technique able to measure  $T_{PR}$ \emph{in
situ}, i.e. in the setup that is used for imaging phantoms (and
breast in future) \cite{Lesaffre_06}.

\subsection{Principle of the measurement of $T_{PR}$ }

\begin{figure}[]
\begin{center}
\includegraphics[width =8.2 cm,keepaspectratio=true]{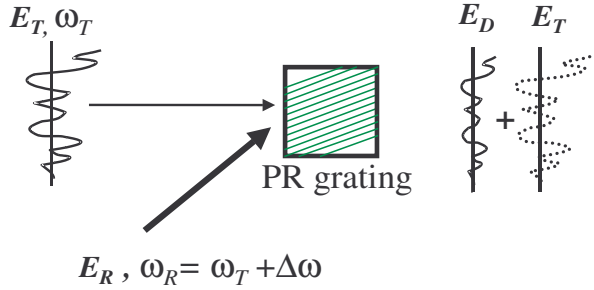}
\caption{PR effect when the selected photons and reference beam
frequencies are different: $\omega_{R} \ne \omega_T$}
\label{fig_meas_TPR}
\end{center}
\end{figure}

To perform acousto-optic imaging with living sample, it is very
important to measure $T_{PR}$, since it is necessary to match
$T_{PR}$ with the sample decorrelation time in order to optimize
the detection efficiency. The ability in our setup to freely
choose the frequency of the reference beam gives new opportunity
to measure the photorefractive time $T_{PR}$.

The idea of the measurement technique is illustrated by
Fig.\ref{fig_meas_TPR}. The frequency of the reference beam
$\omega_R$ is shifted with respect to the signal beam frequency
$\omega_T=\omega_L+ \omega_A$ or $\omega_L$:
\begin{equation}\label{Eq_ER_frequency}
    \omega_R=\omega_T+\Delta \omega
\end{equation}
To simplify the notation, we will consider that $E_R$ still
represents a complex field projection at frequency $\omega_T$, so
that $E_R$ must be replaced in the calculation by $E_R~e^{j\Delta
\omega t}$. We get thus:
\begin{equation}\label{Eq_delta_n_bis}
    \delta n(t)  \propto \frac{ \left< E_T E^*_R ~e^{-j\Delta
\omega t} \right>_{T_{PR}}}{|E_T|^2 + |E_R|^2}
\end{equation}
and
\begin{equation}\label{Eq_E_D_bis}
    E_D(t) \propto \frac{E_R
~e^{j\Delta \omega t}}{T_{PR}} ~\frac{ \int_0^\infty  E_T(t-\tau)~
E^*_R ~e^{-j\Delta \omega \tau}~  e^{-\tau/T_{PR}}~ d\tau}{|E_T|^2
+ |E_R|^2}
\end{equation}

Note that if $\Delta \omega $ is zero, Eq.\ref{Eq_E_D_bis} is
identical to Eq.\ref{Eq_E_D}. Note also that $E^*_R$ does not
depend on time, and can thus be removed from the integral.

Consider that a PR experiment is made with a sample whose
decorrelation time is much longer than $T_{PR}$. This means that
decorrelation can be neglected, and that $E_T(t)$ is uniquely
driven by the US amplitude or phase modulation. $E_T(t)$ is thus
known. $|E_T|^2$ can also be neglected in Eq.\ref{Eq_E_D_bis}
denominator (since $|E_T|^2 \ll |E_R|^2$) and thus : \

\begin{equation}\label{Eq_E_D_bis_bis}
    E_D(t) \propto \frac{~e^{j\Delta \omega t}}{T_{PR}} ~ \int_0^\infty E_T(t-\tau)~
e^{-j\Delta \omega \tau}~ e^{-\tau/T_{PR}}~ d\tau~e^{j\Delta
\omega t}
\end{equation}

We must notice that in Eq.\ref{Eq_E_D_bis_bis}, $E_T(t)$ is
convolved by two time kernels. The first kernel $e^{-\tau/T_{PR}}$
is unknown (since $T_{PR}$ is unknown), while the  second
$e^{-j\Delta \omega \tau}$ is known. Its width can be freely
adjusted by tuning $\Delta \omega$ with the acousto-optic
modulator.

From Eq.\ref{Eq_E_D_bis} it is then  straightforward to calculate
the dependance of the acousto-optic signal with $\Delta \omega$
for the different detection configurations (phase modulation for
the tagged photons and amplitude modulation for the untagged
photons) \cite{Lesaffre_06}. Comparing the calculated spectrum to
the experiment yields then an accurate measurement of $T_{PR}$.

\subsection{Calculation of the tagged photons signal with US amplitude modulation}

It is possible to  calculate the tagged photons signal as a
function of $\Delta \omega$ in the phase modulation configuration.
Nevertheless, as shown in \cite{Lesaffre_06}, the shape of the
spectrum is quite cumbersome, and it seems quite heavy to fit the
experimental data with such a spectrum shape.

It is thus more efficient to measure $T_{PR}$ with  a rectangular
amplitude modulation of the US with $50 \%$ cycling ratio, the
tagged (or untagged) acousto-optic signal being measured with a
lock-in amplifier tuned at the modulation frequency. This is the
key point of the detuning method, since measurements are performed
\emph{in situ} at the US modulation frequency (here $2.5kHz$) and
thus do not depend on the frequency response of the detector,
which is quite distorted due to the many stages of electronic
filters connected to the photodetector. A straightforward
calculation gives only three contributions for the $P$ and $Q$
quadrature of the lock-in signal \cite{Lesaffre_06}  :
\begin{eqnarray}\label{Eq_PQ}
    P=P_0+P_{+}+P_{-}\\
    Q=Q_0+Q_{+}+Q_{-}
\end{eqnarray}
with
\begin{eqnarray}\label{Eq_12a}
    P_{0}(\Delta\omega)=\frac{2A}
{1+(\Delta\omega T_{PR})^{2}}
\end{eqnarray}
\begin{eqnarray}\label{Eq_12b}
    P_{\pm}(\Delta\omega)=\frac{A}{1+(\omega_{mod}\mp\Delta\omega)^{2}~T_{PR}^{2}}
\end{eqnarray}
%
%
\begin{eqnarray}\label{Eq_13}
Q_{\pm}(\Delta\omega)=-\frac{A(\omega_{mod}\mp\Delta\omega)T_{PR}}
{1+(\omega_{mod}\mp\Delta\omega)^{2}~T_{PR}^{2}}
\end{eqnarray}
where $A$ is a proportional constant. By using Eq.\ref{Eq_12a} to
Eq.\ref{Eq_13} it is then quite simple to measure $T_{PR}$ by
fitting the experimental data with the calculated $\Delta \omega $
spectrum.

\subsection{Measurement with the tagged photons and US amplitude modulation}

\begin{figure}[]
\begin{center}
\includegraphics[width =8.2 cm,keepaspectratio=true]{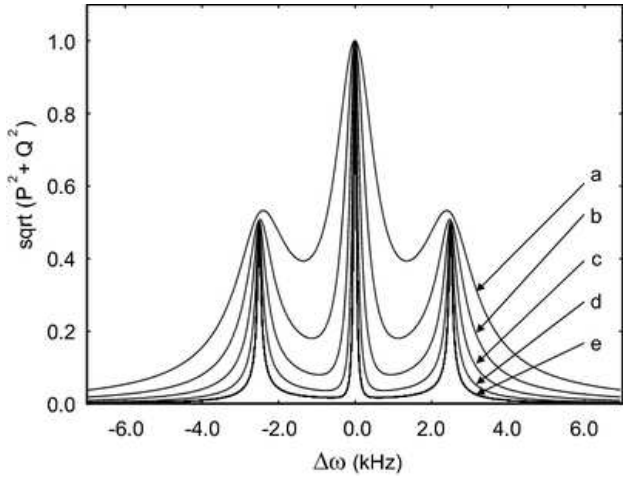}
\caption{Calculated spectrum of the tagged photons signal
(amplitude modulation with $50 \%$ cycling ratio) with
$T_{PR}=0.25ms$ (a), $0.5 ms$, (b), $1ms$ (c), $2ms$ (d), $4ms$
(e). Horizontal axis is the frequency offset $\Delta \omega$.
Vertical axis is $\sqrt{P^2 + Q^2}$ lock-in signal in arbitrary
normalized units.} \label{fig_curve_mod_ampl_simu}
\end{center}
\end{figure}

From Eq.\ref{Eq_PQ} to Eq.\ref{Eq_13} we have calculated the
tagged photons signal as a function of $\Delta \omega$, when the
US beam is modulated with a rectangular $[0,1]$ amplitude
modulation of $50 \%$ cycling ratio, the detection being performed
with a lock amplifier tuned at the modulation frequency
($2.5kHz$). As seen on Fig.\ref{fig_curve_mod_ampl_simu} the shape
of the spectrum is strongly dependent on $T_{PR}$.

\begin{figure}[]
\begin{center}
\includegraphics[width =8.2 cm,keepaspectratio=true]{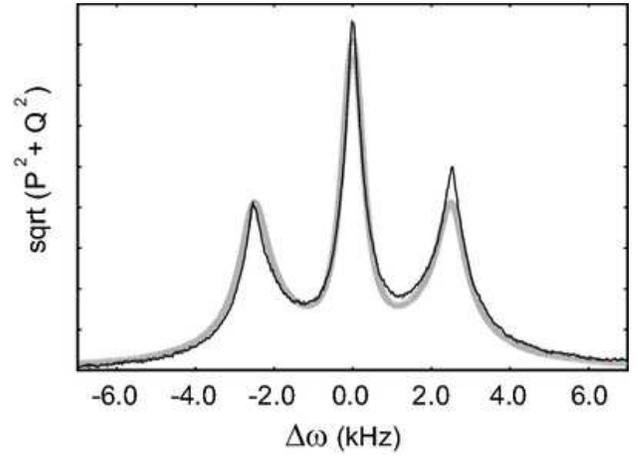}
\caption{Spectrum of the tagged photons signal  with $50\%$ duty
cycle amplitude modulation of the US. Black curve is experimental
data with reference beam flux of $300 mW/cm^2$ and a modulation
frequency of $2.5kHz$. Heavy grey line curve is calculated
spectrum  with $T_{PR}=0.45ms$. Horizontal axis is the frequency
offset $\Delta \omega$.  Vertical axis is $\sqrt{P^2 + Q^2}$
lock-in signal in arbitrary  units.}
\label{fig_curve_mod_ampl_untag}
\end{center}
\end{figure}

It is then possible to fit experimental data on the theoretical
curves. Fig.\ref{fig_curve_mod_ampl_untag} shows the magnitude
(\emph{e.g} $R=\sqrt{P^2 + Q^2}$) of the lock-in signal (points).
The tagged photons are selected ($\omega_R = \omega_L + \Delta
\omega$) and the US beam is modulated in amplitude with $50\%$
duty cycle. The reference beam flux of $300 mW/cm^2$, and  the
modulation frequency is $2.5kHz$. We have fit the experimental
data  with the theoretical curve deduced from Eq.\ref{Eq_12a} to
Eq.\ref{Eq_13}. The fit free parameters are $T_{PR}$ and $A$. The
best fit yields $T_{PR}=0.45ms$. The experimental data are on
Fig.\ref{fig_curve_mod_ampl_untag} as a black curve, the fit as an
heavy grey line curve.

\begin{figure}[]
\begin{center}
\includegraphics[width =8.2 cm,keepaspectratio=true]{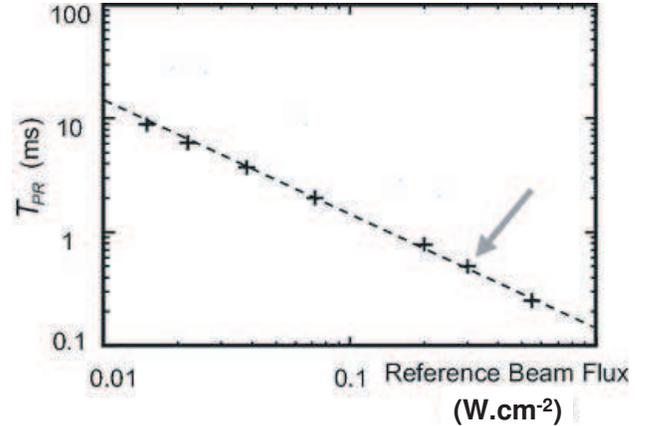}
\caption{$T_{PR}$ in $ms$ as a function of the  reference beam
flux in $W/cm^2$. Crosses are experimental data. Dashed line is
data extrapolation with slope $-1$. Grey arrow corresponds to
Fig.\ref{fig_mod_phase__signal} experimental conditions.}
\label{fig_TPR_I}
\end{center}
\end{figure}

We have recorded many spectra for different reference beam flux.
Each spectrum has been fitted by the theoretical curve yielding
$T_{PR}$. Fig.\ref{fig_TPR_I} shows in log-log scale $T_{PR}$ as a
function of the  reference beam flux. Experimental points are
crosses, data linear log-log extrapolation is dashed line. As
seen, the slope of the extrapolation line is $-1$. This means that
$T_{PR}$ is inversely proportional to the beam flux, as expected.

The shortest photorefractive time we get is $T_{PR}=0.25ms$ for a
flux of 0.55 W/cm$^2$. The Fig.\ref{fig_mod_phase__signal}
modulation phase signal is obtained with $T_{PR}=0.5ms$ and
$0.3W/cm^2$ (grey arrow on Fig.\ref{fig_mod_phase__signal}).

\section{Conclusion}

Seeing through highly scattering media such as living  tissues is
a goal difficult to reach. Coupling light and ultrasound in
acousto-optic imaging is a promising method to reach this aim.
Nevertheless the efficient detection of the tagged photons remains
a challenge.

The PR crystal detection scheme proposed here is a possible way to
solve this problem. PR crystal detection has many advantages. The
detection optical \textrm{etendue} is large since photodetector
area may be quite large ($\sim 1 \textrm{cm}^2$), and since the
collecting lens numerical aperture can be large, too (NA $\sim 1
$). Since the detector is a single-detector (photodiode), the
analysis of the data is simple and fast. By adjusting the power of
the pump beam, it is possible to match the detection bandwidth
$1/T_{PR}$ with the signal bandwidth $\Delta \omega$ in order to
detect with optimal efficiency the "tagged" or "untagged" signal
diffused by living tissues that are broadened by the diffuser
inner motion (brownian motion, blood flow ...). We demonstrate
here our ability to get a high SNR (see
Fig.\ref{fig_mod_phase__signal}) with a thick chicken sample. Our
chicken sample does not decorrelate as do living tissues, the
measurement is done with a short photorefractive time $T_{PR}=0.5$
ms. This result is very encouraging.

The results presented in this paper have been obtained with a
Nd:YAG laser at 1064 nm and a GaAs photorefractive crystal. The
method could be significantly enhanced by the use of a laser
source at 800 nm, according to the absorption coefficients of
hemoglobin and de-oxyhemoglobin, in order to perform a measurement
of the local blood activity (two wavelengths measurements). We are
searching at present for new PR crystals that are sensitive in
this spectral range.

This work is currently supported by a grant from the project
Canc\'eropôle Ile-de-France.\\



\end{document}